\documentclass[english,nofootinbib,notitlepage
]{revtex4}

\usepackage[T1]{fontenc}
\usepackage[latin9]{inputenc}
\usepackage{color}
\usepackage{graphicx}
\usepackage{amssymb,amsmath}
\usepackage{babel}

\def\be{\begin{equation}}
\def\ee{\end{equation}}

\makeatletter

\@ifundefined{definecolor}
 {\usepackage{color}}{}

\begin{document}

\title{The Higgs mass range from Standard Model false vacuum Inflation  \\ in scalar-tensor gravity}

\author{Isabella Masina$^{1,2}$}
\email{masina@fe.infn.it}
\author{Alessio Notari$^{3}$}
\email{notari@ffn.ub.es}

\affiliation{$^{1}$  Dip.~di Fisica, Universit\`a di Ferrara and INFN Sez.~di Ferrara, Via Saragat 1, I-44100 Ferrara, Italy}
\affiliation{$^{2}$ CP$^3$-Origins \& DIAS, Southern Denmark University, Campusvej 55, DK-5230 Odense M, Denmark}
\affiliation{$^{3}$ Departament de F\'isica Fondamental i Institut de Ci\`encies del Cosmos, Universitat de Barcelona, Mart\'i i Franqu\`es 1, 08028 Barcelona, Spain}

\begin{abstract}
If the Standard Model is valid up to very high energies it is known that the Higgs potential can develop a local minimum 
at field values around $10^{15}-10^{17}$ GeV, for a narrow band of values of the top quark and Higgs masses.
We show that in a scalar-tensor theory of gravity such Higgs false vacuum can give rise to viable inflation  
if the potential barrier is very shallow, allowing for tunneling and relaxation into the electroweak scale true vacuum.
The amplitude  of cosmological density perturbations from inflation is directly linked to the value of the Higgs potential at the false minimum.
Requiring the top quark mass, the amplitude and spectral index of density perturbations to be compatible with observations, 
selects a narrow range of values for the Higgs mass, $m_H=126.0\pm 3.5$ GeV, where the error is mostly due to the theoretical
uncertainty of the 2-loop RGE. 
This prediction could be soon tested at the Large Hadron Collider. 
Our inflationary scenario could also be further checked by better constraining the spectral index and the tensor-to-scalar ratio.

\end{abstract}


\maketitle


\section{Introduction}

Despite the many experimental successes of the Standard Model (SM) of particle physics, its scalar sector is still to be confirmed by accelerator experiments. 
Extensions of the SM are intensively explored, motivated by the fact that the SM does not explain
neutrino masses, dark matter, the baryon-antibaryon asymmetry, a primordial stage of inflation, etc.
From a theoretical  point of view,  useful guides for extending the SM are represented by a possible unification of gauge couplings,
by fine-tuning problems (such as the lightness of the Higgs mass) and by the inclusion of gravity.

We consider here the possibility that no major extensions of the SM Higgs sector are required. 
In this framework, we consider the possibility of realizing inflation by introducing only modifications in the gravitational sector.

The idea that inflation is realized in the SM Higgs sector has been proposed recently ~\cite{arXiv:0710.3755}, 
by allowing for a non-minimal coupling of the Higgs to gravity. 
This is an appealing point of view because it does not introduce new degrees of freedom. 
It is however controversial whether such scenario is viable at the quantum level, because radiative corrections could modify the potential
\cite{arXiv:0809.2104,arXiv:0812.4946}.

In this paper we propose a slightly less minimalistic scenario, in which we {\it do} introduce a new scalar degree of freedom, but assume it to be decoupled 
from the SM: in such a way the SM is not altered at all and only gravitational physics is affected.

We use the fact that the Higgs potential has a local minimum between $10^{15}-10^{17}$ GeV, which is indeed known to exist for a narrow 
band of the top and Higgs mass values \cite{CERN-TH-2683,hep-ph/0104016,strumia2}.
We will assume that the Universe started with the Higgs in this false vacuum, which leads to a stage of exponential inflation, where density
perturbations are produced. 

It is already a non-trivial fact that such a local minimum can exist in the Higgs potential, but is even more non-trivial that there exist an
allowed parameter range in the Higgs and top masses, for which the energy density of this minimum has the right value (GUT scale)
to give rise to the correct amplitude of density perturbations from inflation.

The additional scalar has the  role of providing a {\it graceful exit} to inflation, which is possible in a  Brans-Dicke scalar-tensor theory of
gravity, as shown in ref.~\cite{astro-ph/0511396,hep-ph/0511207} (and earlier in ~\cite{johri,extended,hyperextended} for power-law inflation). 
After the stage of exponential inflation,  the expansion is drastically slowed down by the Brans-Dicke scalar and the Higgs field can tunnel
out efficiently, via production and collision of bubbles, and subsequently rolls down to its present vacuum expectation value, $v \sim 246$ GeV.
Moreover quantum fluctuations in the Brans-Dicke scalar generate a spectrum of density perturbations, which can be understood also as a tunneling which happens at slightly different times in different regions of space.
In particular, we show that successful inflation can be achieved for a very broad class of couplings of the scalar-tensor theory 
and that the spectral index of density perturbations is largely independent on the particular choice of couplings.

We use the information from the amplitude and spectral index of cosmological density perturbations generated during inflation to predict the Higgs mass, finding
\be
m_H= (126.0 \pm 3.5) \,{\rm GeV} \,\,.
\ee
where the error is mostly due to the theoretical uncertainty of the 2-loop RGE. 
This range is within the present experimental window set by direct searches, namely the $115$ GeV lower bound set by LEP \cite{PDG} 
and the $140$ GeV  upper bound set recently by LHC \cite{HCP11}, which restricted the previous $155$ GeV bound set by Tevatron \cite{arXiv:1107.5518}.
We note that the range above is also compatible with the July 2011 global electroweak SM precision fit \cite{GFitter},
which gave $m_H=125^{+8}_{-10} $ GeV at $1\sigma$.

After the first release of this paper on the arXiv, LHC updated the Higgs boson exclusion limits \cite{last}:
as of 13 December 2011 ATLAS excludes at the $95 \%$ CL masses outside $116-130$ GeV and CMS excludes at the $95\%$ CL masses 
outside $115-127$ GeV.
Remarkably, the allowed band for SM false vacuum inflation includes the region $124-127$ GeV, 
where ATLAS and CMS recorded excesses of events in the di-photon as well as 4-leptons channels \cite{last}. 
At present, the significance of this result is however still low.

Hopefully, our inflationary model prediction will be soon better tested by LHC.
In the case of compatibility with LHC results, future more precise cosmological measurements of the scalar spectral index and the tensor-to-scalar ratio
could provide further tests for the idea of Higgs false vacuum inflation \cite{Masina:2011un}.

The paper is organized as follows. In section \ref{Model} we present our inflationary model, based on a scalar-tensor theory of gravity.
First, we derive the expression for the slow roll parameters, the scalar spectral index and the tensor-to-scalar ratio.
We then establish the relation between the amplitude of density perturbations and the Higgs potential at the false minimum. 
In section \ref{Minimum} we study values of the top and Higgs masses giving a SM false vacuum within the range consistent with inflation.
In section \ref{Results} we show that the set of values derived in this way are only in part consistent with the measurement of $m_t$.
This allows to predict a narrow range for the Higgs mass.  In section \ref{Post} we provide details about the post-inflationary period, while in 
section \ref{General} we generalize the model by considering higher order terms.
We finally draw our conclusions in section \ref{Conclusions}.


\section{The inflationary model}
\label{Model}

The model setup is the SM of particle physics in a scalar-tensor theory of gravity. We denote the Higgs field by $\chi$. 
For very large field values, the quadratic term $m^2 \chi^2$ can be neglected in the Higgs potential which, at some scale $\mu\sim \chi$
becomes simply
\begin{equation}
V(\chi) \simeq \lambda(\mu)\chi^4\,\,.
\end{equation}

It is well known that the SM Higgs potential has a false minimum for some narrow band of the Higgs and top masses 
\cite{CERN-TH-2683, hep-ph/0104016, strumia2}, as we also discuss in sections \ref{Minimum} and \ref{Results}.
For such band, the coupling $\lambda$ goes very close to zero at large field values and then rises again, 
namely the Higgs potential can develop a new local minimum, which turns out to be well compatible with the GUT scale range, $10^{15}-10^{17}$ GeV. 
We call $\chi_0$ the Higgs field at the false minimum\footnote{There is also an intermediate regime in which the Higgs can develop a flat region of the potential, 
but it has been shown that inflation is not viable~\cite{strumia2}, at least in standard Einstein gravity.}.

We are going to use this new minimum to drive cosmic inflation, since $V(\chi_0)$ is a large potential energy which can be a source for an exponential
expansion of the Universe. The non-trivial ingredient is to provide a {\it graceful exit} from inflation, that is  a transition to a radiation
dominated era, in a nearly flat Universe, at a sufficiently high-temperature. 
This is known to be impossible in standard gravity~\cite{Guth}. In fact, in order to end inflation the field would have to tunnel to the other side of the potential barrier 
by nucleating bubbles with a different value of the field~\cite{Coleman} and the bubbles would eventually collide with each other and reheat the Universe.
However,  the nucleation rate per unit time and volume $\Gamma$ (which has mass dimension four) has to be suppressed as compared to the 
fourth power of the Hubble rate $H_I$, 
otherwise the Universe would tunnel quickly in a few Hubble times, without providing sufficient inflation.
On the other hand, if $\Gamma\ll H_I^4$ the probability of tunneling is very small and the process is not efficient enough to produce a sufficient number of 
bubbles inside a Hubble horizon, which could collide producing a homogeneous radiation thermal bath.
The possibility that we lived inside one single bubble, without collision with other bubbles, is also ruled out by the fact that the
inner region of a bubble would have too large spatial curvature.
In other words, a graceful exit  would require $\Gamma$ to become larger than $H_I^4$ only after some time, 
but this is  impossible because both quantities are time-independent.

Interestingly, such a behavior becomes possible in a scalar-tensor theory of gravity.  This has been shown in earlier models, under the name of 
extended~\cite{johri,extended} or hyperextended~\cite{hyperextended} inflation and more recently in~\cite{astro-ph/0511396,hep-ph/0511207}.
In earlier models only a power-law inflation (scale factor growing with time as $a(t)\propto t^{\alpha}$) was proposed, but subsequently this has been shown to be in tension with observations of the Cosmic Microwave Background, because it turns out to be difficult to produce a nearly flat spectrum of perturbations~\cite{Liddle}. 
In~\cite{hep-ph/0511207,astro-ph/0511396} it was shown instead that a stage of exponential expansion is naturally incorporated in the model and then followed by a stage of
power-law (even decelerated) expansion. 
In this way, it is possible to produce the correct spectrum of perturbations in the first stage of inflation and subsequently to slow down dramatically the expansion of the Universe, 
thereby allowing the field trapped in the false minimum to tunnel through percolation of bubbles.

A scalar-tensor theory of gravity is obtained by adding a scalar field $\phi$ (sometimes called Brans-Dicke scalar or dilaton), coupled to the Ricci scalar $R$
via a non-minimal coupling. We will follow here very closely the scenario presented in~\cite{astro-ph/0511396,hep-ph/0511207}. 
The full action of the model\footnote{A potential term $U(\phi)$ for the field $\phi$ could be added, and in fact it is likely to be needed for the post-inflationary
evolution. We assume that this potential term is subdominant before the tunneling. } 
is given by
\begin{equation}
-S=\int d^4 x \sqrt{-g} \left[ {\cal L}_{SM} + \frac{(\partial_\mu\phi\partial^\mu\phi)}{2} -\frac{M^2}{2} f(\phi) R \right]  
\label{azione}
\end{equation}
where $ {\cal L}_{SM} $ includes all the SM of particle physics and where we require that $f(\phi)$ is always positive.
Here, for small $\phi$ values we assume that we can simply expand the function $f(\phi)$ as:
\begin{equation}
f( \phi)\simeq 1 +\beta \left( \frac{ \phi}{M} \right)^2  +\gamma \left( \frac{ \phi}{M} \right)^4 + ... \,\,,
\label{eq-f}
\end{equation}
where $M$ plays the role of the Planck mass and $\beta$, $\gamma$ are dimensionless couplings. 
Indeed, the usual case of standard Einstein gravity corresponds to $f=1$.
In general we may view these models just as gravitational theories in which the effective Planck mass -- given by $\sqrt{f} M$  -- is fixed by the vacuum 
expectation value of $\phi$.
For small values of $\phi$ one then recovers the usual Einstein gravity with a Planck mass given by $M$.
As in~\cite{hep-ph/0511207}, for large field values ($\phi\gg M$) we require $f$ to be a monotonic growing function, such that $f(\phi)>\beta (\phi/M)^2$.

We start by assuming that the primordial Universe is initially in a cold state,
where $\phi$ takes a very small value and the Higgs field is trapped in the false minimum of its potential, at a field value $\chi_0$.
The evolution equation~\cite{astro-ph/0511396, espositofarese} of the metric in an FLRW background is given by
\begin{eqnarray}
H^2 &=& \frac{1}{3M^2 f(\phi) }\left[ \frac{1}{2} \dot\phi^2 - 3 H M^2\dot{f}(\phi) + V(\chi_0) \right] \, ,  \label{eqH}
\end{eqnarray}
where $H\equiv\dot{a}/a$ ($a$ is the scale factor) and the evolution equation for the scalar field is:
\be
 \ddot\phi + 3H \dot\phi = \frac{M^2}{2}\frac{df(\phi)}{d\phi} R\, .  \label{eqfi}
\ee

Since $\phi$ is small we can approximate (\ref{eqH}) with the standard equation 
\be
H^2\simeq \frac{V(\chi_0)}{3 M^2} \equiv H_I^2\, ,
\ee
leading to a stage of inflation
with the Universe expanding almost exponentially with a scale factor $a(t)\propto e^{H_I t}$.

In this minimum the Ricci scalar has a value:
\be
R=6 \dot{H}+12 H^2 \simeq 12 H_I^2 \, , \label{Ricci}
\ee
where the second equality follows from the fact that during inflation $H$ is
almost constant.

In this false minimum the Higgs field can tunnel with a Coleman instanton~\cite{Coleman}, a {\it bounce} solution of the classical equations of motion, but it will do it with negligible probability if $\Gamma\ll H_I^4$~\cite{astro-ph/0511396}.
However in such an inflationary background there is a time-dependent quantity, which is the value of the Brans-Dicke field $\phi$, which sets dynamically the
value for the Planck mass.
It is easy to see combining eq.~(\ref{eqfi}) and eq.~(\ref{Ricci}) that, if the function $f(\phi)$ is a monotonic increasing function,
the presence of a nonzero background value for $R$ makes the additional field $\phi$ grow and at some point the field $\phi$ will
reach values large enough so that the Planck mass, given by $f(\phi)$, starts becoming larger than $M$.
When we enter this regime, gravity becomes weaker and so the Hubble parameter starts decreasing with time. If we wait for a sufficiently long
time, the tunneling via a Coleman transition
will happen successfully when $H^4\simeq \Gamma$, since $\Gamma$ is a constant. Note however that the transition could happen also by quantum fluctuations due to the gravitational background, through the Hawking-Moss instanton, which would make also $\Gamma$ time dependent. We now disregard this possibility here and comment it later in section~\ref{Minimum}.

Such a process can be studied by analyzing directly the above equations of motion, derived by the action~(\ref{azione}). It is however simpler and more
general to make a change of variables, as in~\cite{hep-ph/0511207}, and go in the so-called {\it Einstein frame} (we use the bar to indicate a quantity in this frame), which 
is related to the original frame through the change of variable 
obtained via the conformal transformation of the metric $\bar{g}_{\mu\nu}=f(\phi)g_{\mu\nu}$.
The action in this frame becomes (see~\cite{hep-ph/0511207, espositofarese} for further details)
\be
S_{E}= {1\over 2}\int d^4x\ \sqrt{-\bar{g}}[M^2 \bar{R}-K(\phi)(\bar{\partial}\phi)^2- 2 \bar{{\cal L}}_{SM} ]\,\,\,,\,\,\,K(\phi)\equiv {2f(\phi)+3 M^2 f^{'2}(\phi)\over 2f^2(\phi)}
\label{e-action} \, .
\ee

Let us now focus on the function $f(\phi)$, 
which can generically be written for any value of the field $\phi$ as:
\begin{equation}
f( \phi)\simeq 1 +\beta \left( \frac{ \phi}{M} \right)^2  +\sum_{n \ge 4}  \gamma_n \left( \frac{ \phi}{M} \right)^n \ \,\,.
\label{eq-ftot}
\end{equation}
As shown below, it is required ~\cite{hep-ph/0511207} that for large field values $f$ grows faster than $(\phi/M)^2$. This is achieved for instance if all coefficients 
$\gamma_n$ of the higher dimensional operators are positive numbers. 
In order to keep the analysis simple, it is sufficient to focus on one single operator, and the simplest one is the operator with $n=4$.
As already anticipated in eq.(\ref{eq-f}), from now on we therefore consider only the $n=2$ and $n=4$ terms: this allows to have a model
with only two parameters, $\beta$ and $\gamma$.
We will show in section \ref{General} 
that for {\it any} coupling with $n>2$, the predictions are very similar and converge to a single prediction at large $n$. Note that for this reason in our scenario there is a very wide class of higher-dimensional operators 
which work well, and that the predictions are almost independent on the precise functional form of $f$ inside this class. This is different from the case 
proposed by~\cite{arXiv:0710.3755} where only the quadratic term in $f$ works well (making the scenario 
possibly unstable under quantum corrections, because of the appearance of higher dimensional operators).

In terms of a canonically normalized field $\Phi$ defined through $d\Phi=d\phi \sqrt{K(\phi)}$, the action in eq.(\ref{e-action}),
can be further simplified to:
\be
S_{E}= {1\over 2}\int d^4x\ \sqrt{-\bar{g}}[M^2 \bar{R}-(\bar{\partial}\Phi)^2- 2 \bar{{\cal L}}_{SM} ]
\label{e-action-2} \, .
\ee
Because of the conformal transformation to the Einstein frame, the Higgs potential becomes $ V(\chi)/f(\Phi)^{2}$, so that
the potential energy at the false Higgs minimum gives rise to a potential term for $\Phi$
\be
S_{E}^{vac}=\int d^4x\ \sqrt{-\bar{g}} \bar{V} \,\,\,\,, \,\,\, \bar{V}\equiv  \frac{V(\chi_0)}{f(\Phi)^{2}} \,\,. 
\label{eq-Vb0}
\ee
Now, in this frame, we can discuss in a very general way the dynamics, distinguishing between two stages: small $\Phi$ and large $\Phi$.
At small $\Phi$, considering in (\ref{eq-ftot})  only the $n=2,4$ terms as discussed, and assuming $\beta^2 \ll \gamma$,  the Higgs potential becomes
\be
\bar V\left(\Phi\right) =V(\chi_0) \left(   1 -2 \beta \left( \frac{ \Phi}{M} \right)^2    - 2  \gamma   \left( \frac{ \Phi}{M} \right)^4 + ...  \right)\,\,.
\label{eq-Vb}
\ee
This acts as a hill-top potential for the $\Phi$ field. 
So, in this frame $\Phi$ rolls down the potential from small to high values. If $\beta \ll 1$,  the field rolls down slowly and the standard
slow-roll approximation can be used.

For large field $\Phi$ values, instead, it can be seen that, under the assumption that $f(\phi)$ grows faster than quadratic, we 
have that $M^2 f^{'2}>|f|$.
In this case in the numerator of $K$ in eq.~(\ref{e-action}), the second term dominates.
Therefore, in this phase, the kinetic term can be approximated as
\be 
K(\Phi)\approx {3\over 2}\left(M f'\over f\right)^{2}\, \, .
\ee
We can now write the canonical variable in this regime simply as
\be
\Phi =   \sqrt{3\over 2} M \ln f(\phi) \, ,
\ee
so that the potential for $\Phi$, eq.(\ref{eq-Vb0}), becomes remarkably independent on the exact form of $f$:
\be
\bar{V}(\Phi)=V(\chi_0) \exp\left({-2\sqrt{\frac{2}{3}}{\Phi\over M }}\right) \, .
\label{potential}
\ee
Evolutions under such exponential potential corresponds to a power law phase, with decelerated expansion, given by
$
\bar{a}\sim \bar{t}^{3/4}.
$
It is also easy to see that $\Phi$ grows and the kinetic energy is always proportional to $\bar{V}$ (precisely it is $4/5 \, \bar{V}$).

Now, the end of this phase is achieved when $\bar{H}^2\simeq \frac{\bar{V}}{M^2}$ is equal to $\bar{\Gamma}^{1/2}$ and at this point the Higgs field tunnels efficiently. 
Therefore the final field value at tunneling $\Phi_T$ is given by:
\be
f(\Phi_T)\simeq \frac{V(\chi_0)^{1/2}}{M \bar{\Gamma}^{1/4}}=  \frac{V(\chi_0)}{M^2 {\Gamma}^{1/2}} \label{phif} \, ,
\ee
where the last equality has been derived using the fact that the dimensionful parameter $\Gamma$ rescales between the two frames as  $\bar{\Gamma}=\Gamma/f^2$.
 In principle $\Gamma$ is calculable knowing the SM potential exactly, requiring an accurate numerical solution of the bounce equation~\cite{Coleman}.
However, the quantity $\Gamma$ is exponentially sensitive to the SM parameters, and so we cannot compute  it with the present known 
experimental errors. 
For this reason we will treat it as a free parameter, which leads to the conclusion that also $\Phi_T$ is a free parameter. 
However, for practical purposes we only need to know that the transition is possible, leading to a radiation-dominated Universe. 
Knowing when the transition happens can change only the number of e-folds which correspond to our horizon scale today. 
For any practical purpose  we leave this as a free parameter, as we discuss in section~\ref{Post}, and focus now on the 
observational consequences of the slow-roll stage.

Assuming the initial value for $\Phi$ to be of the order of the quantum fluctuations, given by $\Phi\approx H_I$, 
it is easy to check that the total number of e-folds is always huge.

\subsection{Slow-roll parameters and the number of e-folds}

We are now in the position to calculate the slow roll parameters in our model: 
\be 
\epsilon(\Phi) = \frac{1}{2} \left| \frac{1}{\bar V}  \frac{d \bar V}{d(\Phi/M) }\right|^2 
\approx  8  \left( \frac{ \Phi}{M} \right)^2  \left( \beta + 2 \gamma   \left( \frac{ \Phi}{M} \right)^2  \right)^2\,\,,
\ee
\be
\eta(\Phi)= \frac{1}{\bar V}  \frac{d^2 \bar V}{d(\Phi/M)^2 } \approx -4 \beta - 24 \gamma  \left( \frac{ \Phi}{M} \right)^2 \,\,.
\ee
Inflation ends when one of these parameters becomes of order one. We call $\Phi_f$ the value of the field $\Phi$
at the end of inflation, for definiteness say when $\epsilon$ or $\eta$ become exactly equal to one.
The number of e-folds corresponding to a smaller value of $\Phi$ is
\be
N(\Phi) \approx \frac{1}{M} \int_{\Phi}^{\Phi_f} d{\Phi'}\, \frac{1}{\sqrt{2 \, \epsilon(\Phi')}}
\approx \frac{1}{8 \beta} \ln \frac{ \left( \frac{ \Phi}{M} \right)^2 2 \gamma +\beta}{\left( \frac{ \Phi}{M} \right)^2 2 \gamma + \left(\frac{\Phi}{\Phi_f}\right)^2 \beta}\,\,.
\label{Nphi}
\ee  

For what concerns the small $\phi$ regime both couplings $\beta$ and $\gamma$ could be important, 
but in order to have the simplest possible model we now focus on the case $\beta=0$ (we will analyze in more detail the case $\beta\neq0$ in section
\ref{General}); 
this model has just one parameter, $\gamma$, in addition to those of the SM. The slow roll parameters become simply:
\be
\epsilon(\Phi) \approx  32 \gamma^2 \left( \frac{ \Phi}{M} \right)^6 \,\,\,\,\,,\,\,\,\,\, \eta(\Phi) \approx  -24 \gamma  \left( \frac{ \Phi}{M} \right)^2 \,\,.
\label{eq-ee-b0}
\ee
For $\gamma \lesssim 2 \times 10^{-3}$, $\epsilon$ becomes of order one  at $(\Phi_f/M)^2 \approx 1/(32 \gamma^2)^{1/3}$, before this happens to $\eta$. 
On the contrary, for $\gamma \gtrsim 2 \times 10^{-3}$, $\eta$ becomes of order one at $(\Phi_f/M)^2 \approx 1/(24 \gamma)$, before $\epsilon$.
The relation between $\Phi$ and the associated number of e-folds becomes:
\be
\left( \frac{ \Phi}{M} \right)^2 \approx \frac{1}{ 16 \gamma N(\Phi) + C_\gamma } \,\,
\label{eq-Phi}
\ee
where $C_\gamma \approx (32 \gamma^2)^{1/3}$ for $ \gamma \lesssim 2 \times 10^{-3}$, while $C_\gamma \approx 24 \gamma $ for $\gamma \gtrsim  2 \times 10^{-3}$.
The expression above allows to calculate the slow roll parameters of eq. (\ref{eq-ee-b0}) as functions of the number of e-folds.

In the case in which $\beta$ is non-vanishing but such that $\beta \lesssim 5\times 10^{-4}$, it turns 
out (see section \ref{General}) that the expressions above specific to the case $\beta=0$
are marginally affected. This means that all the results that we are going to derive in the next section putting formally $\beta=0$
are slightly more general.


\subsection{Tensor to scalar ratio and scalar spectral index}

We call $\bar N$ the number of e-folds corresponding to the present horizon of $3000/h$ Mpc, 
which is expected to be in the range $40\lesssim \bar N \lesssim 60$, as discussed in section \ref{Post}.

We are interested in $\epsilon_{\bar N}$, $\eta_{\bar N}$, since the observable quantities \cite{Peiris:2003ff} are 
the tensor to scalar ratio $r \equiv P_T/P_S = 16 \epsilon_{\bar N}$ and the 
scalar spectral index $n_S=1 -6 \epsilon_{\bar N} +2 \eta_{\bar N}$.
When  $N(\Phi)$  in eq.(\ref{eq-Phi}) is taken to be as large as $\bar N$, the term with $C_\gamma$ in the denominator 
of the eq.(\ref{eq-Phi}) is negligible for both $\gamma$ regimes provided $\gamma \ge 10^{-5}$, in which case we obtain:
\be
n_S \approx  1 - \frac{3}{\bar N}\,\,\,\,,\,\,\,\,\,\,\,r \approx \frac{1}{8 \gamma {\bar N}^3} \,\,\,\,.
\label{eq-rns}
\ee

\begin{figure}[th!]
\includegraphics[width=7cm]{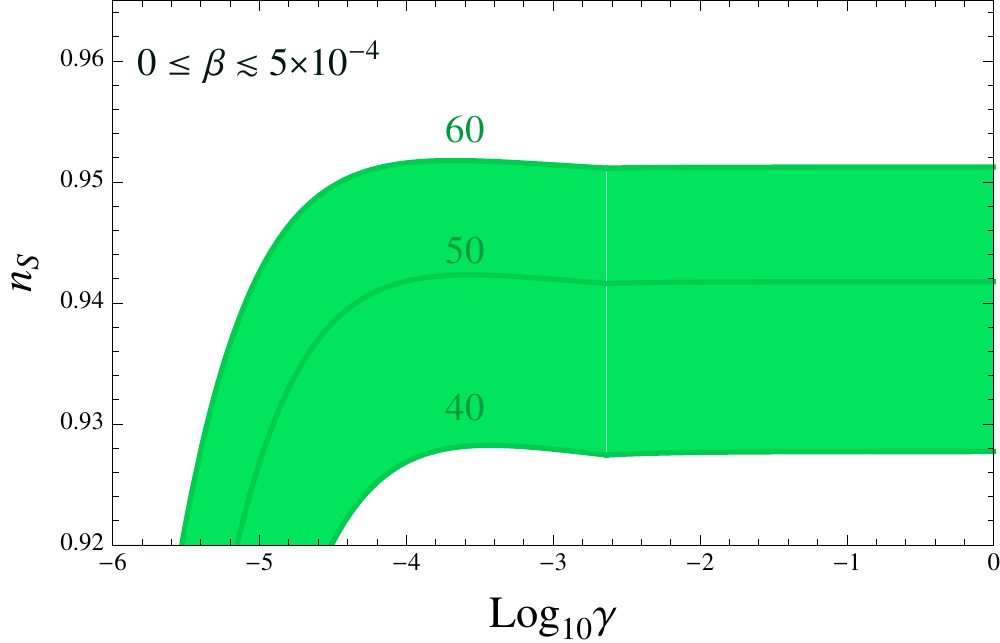} \,\,\,\,\, \,\, 
\includegraphics[width=7cm]{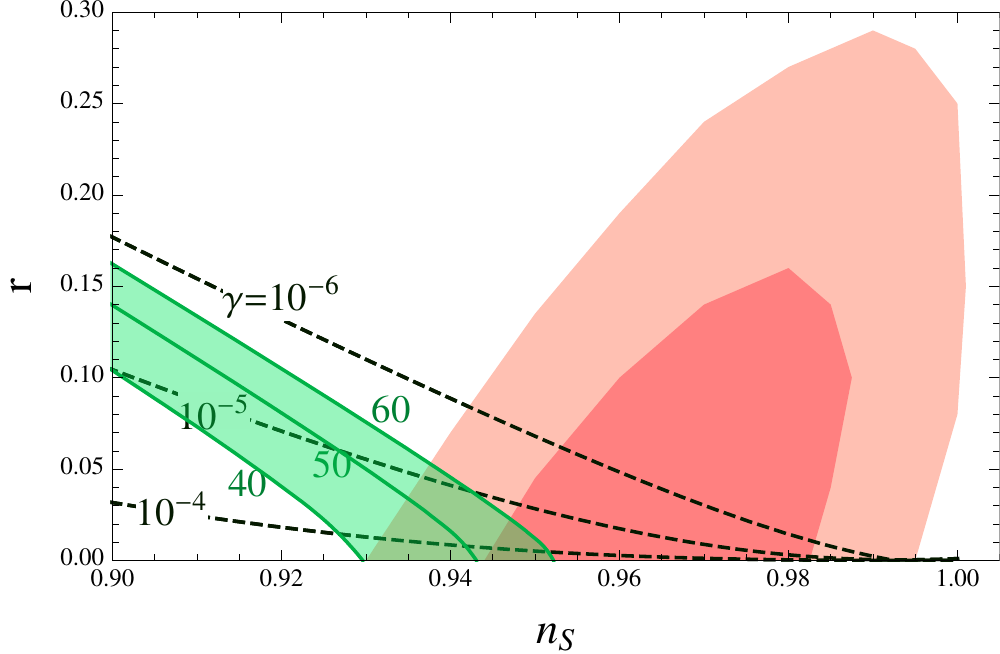}
\caption{ Left: dependence of $n_S$ on $\gamma$ when $0\le \beta \lesssim 5 \times 10^{-4}$. 
The curves are obtained for the representative values of $\bar N=40,50,60$.
Right: dashed curves display $r$ as a function of $n_S$ for selected values of $\gamma$ (and taking $\beta=0$). 
The narrow shaded (green) region is found by requiring $\bar N \approx 50 \pm 10$. 
The larger shaded (red) regions are to the $1$ and $2\sigma$ ranges allowed experimentally for $r$ and $n_S$ \cite{Komatsu:2010fb}. }
\label{fig-nsg-b0}
\end{figure}

The fact that $n_S$ mildly depends on $\gamma$ provides  $\gamma \ge 10^{-5}$ is shown in the left panel of fig.\ref{fig-nsg-b0}. 
Considering for example $\bar N= 50 \pm 10$, we obtain $n_S=0.94 \pm 0.01$, precisely inside its $2\sigma$ experimentally preferred region \cite{Komatsu:2010fb},
as depicted in the right panel.
Clearly, future experiments  with a better precision on $n_S$ could further check this model. 
As shown in section \ref{General},  if we consider in eq.(\ref{eq-ftot}) a term with $n>4$, the prediction for $n_S$ leads to a slightly higher value
 which goes closer to the central measured value, $n_S \approx 0.96$.

The prediction for $r$ depends strongly on $\gamma$ and, as will be discussed in more detail in section \ref{General}, 
very mildly on $\beta$ as far as $\beta \lesssim 10^{-2}$. 
To make the connection with experiment even more direct,  in the right panel of fig.\ref{fig-nsg-b0}
we show $r$ as a function of $n_S$ for various values of $\gamma$ and taking $\beta=0$. The narrow shaded region is found by requiring $\bar N$
to be in the range of interest, $\bar N = 50 \pm 10$. One then realizes that for such range, only values of $\gamma$ larger than $10^{-5}$
allow both $r$ and $n_S$ to be in their experimentally preferred region \cite{Komatsu:2010fb}.
From now on we will therefore consider the parameter $\gamma$ of our inflationary model to be in the range $10^{-5}\le \gamma \le 1$.

\subsection{Amplitude of perturbations and Higgs potential at the false minimum}

The amplitude of density perturbations in $k$-space is specified by the power spectrum:
\be
P_s(k)=\Delta_R^2 \left( \frac{k}{k_0} \right)^{n_S-1}
\ee
where $\Delta_R^2$ is the amplitude at some pivot point $k_0$, predicted by inflation to be
\be
\Delta_R^2 = \left. \frac{\bar V}{ 24 \pi^2 M^4 \epsilon } \right|_{k_0}\,\,.
\label{eq-Dr}
\ee

According to eq.(\ref{eq-Vb}), since we are in the small field regime for $\Phi$, the Higgs potential at the false minimum can be related to $r$ of eq.(\ref{eq-rns}) 
by
\be
\frac{V(\chi_0)}{M^4} = \frac{3}{2} \pi^2 \Delta_R^2  \, r    \,\,.
\ee
 In the following we consider $\Delta_R^2= (2.43 \pm 0.11)\times  10^{-9}$,  taken as the best-fit value from \cite{Komatsu:2010fb}
 for a pivot scale $k_0=0.002 {\rm Mpc}^{-1}$.

\begin{figure}[h!]
\includegraphics[width=9 cm]{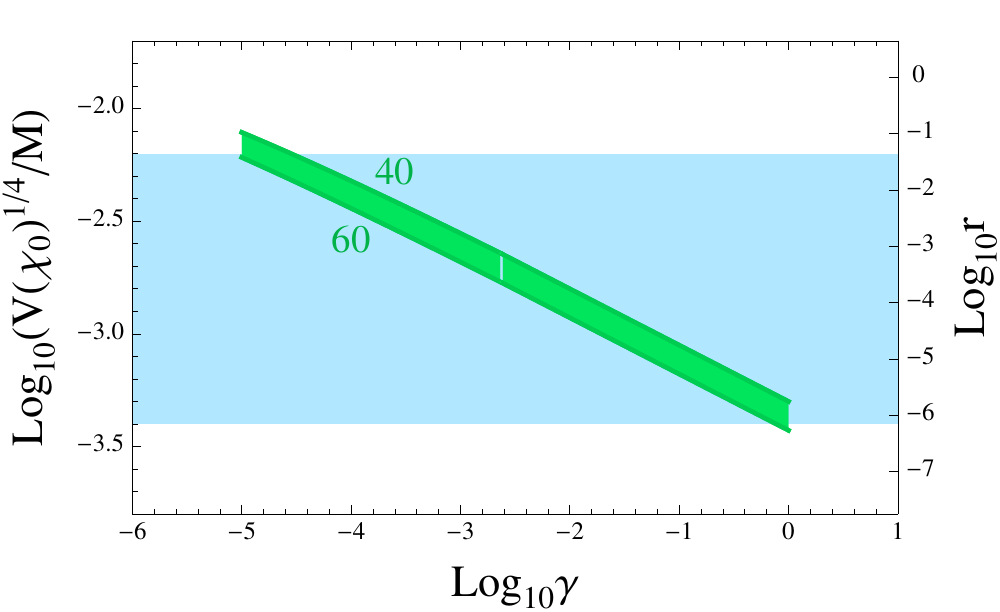} 
\caption{Value of the Higgs potential at the false minimum as a function of $\gamma$ and for $40\le \bar N \le 60$. The shaded region shows that the 
range selected by our inflationary model is $10^{-3.4} \le V(\chi_0)^{1/4}/M\le 10^{-2.2}$.}
\label{fig-Vinf}
\vskip .2cm
\end{figure}

The value of the Higgs potential at the false minimum is shown in fig. \ref{fig-Vinf} as a function of $\gamma$ and for $40 \le \bar N \le 60$.
This plot gives the window of values of the Higgs potential in the false minimum that are compatible with our inflationary model. 
Taking $M= (8\pi G_N)^{-1/2}=1.22 \times 10^{19} /\sqrt{8 \pi}$, where  $G_N$ is the Newton constant, we obtain
\be
9.7\times 10^{14}  \,{\rm GeV}< V(\chi_0)^{1/4}  <  1.5 \times 10^{16}\, {\rm GeV}\,\,,
\ee 
which is, by the way, the range where unification of couplings might take place. 
As we are going to discuss, in the SM this range of values for the Higgs potential at a false minimum $\chi_0$ is natural,
even though it requires highly correlated values for the top and Higgs  masses.

\section{Finding a viable minimum}
\label{Minimum}

The fact that the Higgs field in the SM can develop a false minimum is non-trivial by itself and, as we are going to discuss,
it is even more intriguing that this happens at the right energy scales required by our inflationary model. 

The false minimum requires very specific values of the top and Higgs masses. 
Using 2-loop RGE and matching conditions as discussed {\it e.g.} in \cite{hep-ph/0104016}, we studied such values.  
Of course, the extremely precise values for $m_t$ and $m_H$ that we are going to present are not to be taken sharply, because of a theoretical 
uncertainty of about $3$  GeV on the Higgs mass and about $1$ GeV on the top mass, which is intrinsic in the 2-loop RGE running procedure 
(more on this later). 

\begin{figure}[h!]
 \includegraphics[width=6.7cm]{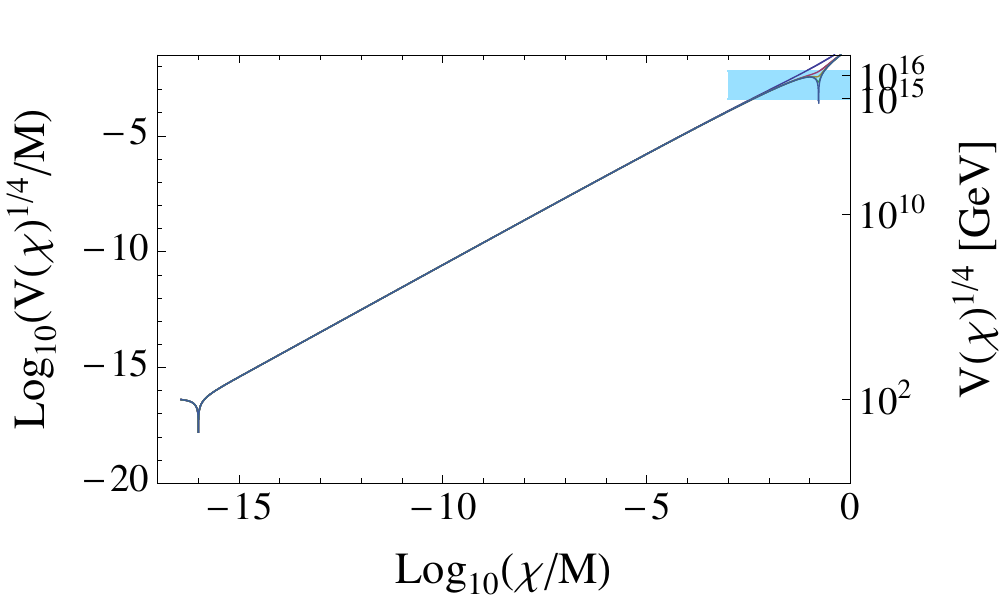} \,\,\,\,\,\,\,\,\quad  \,\,\,\,\,\, \includegraphics[width=7.1cm]{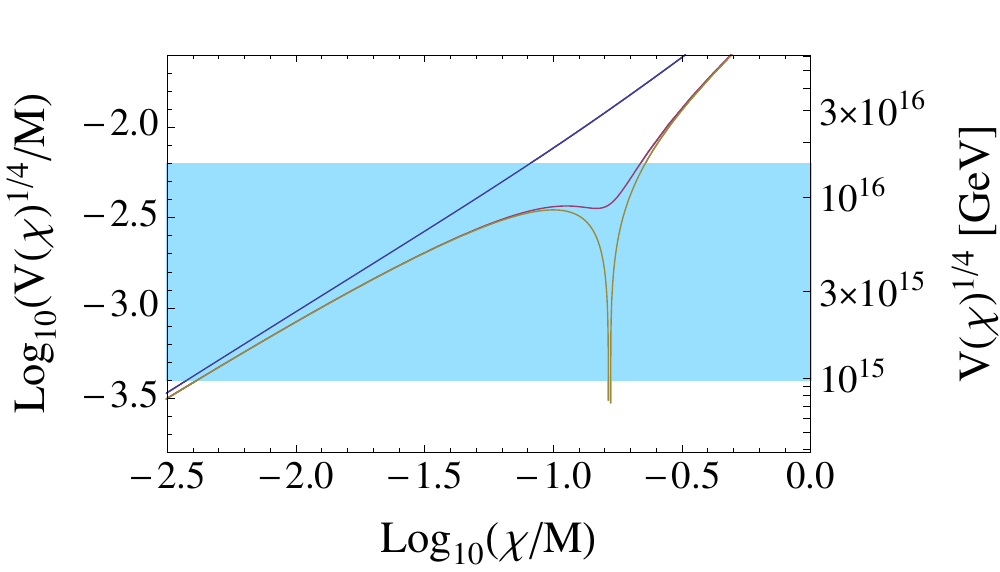}
\caption{Higgs potential as a function of the Higgs field value $\chi$. We fixed $m_t=171.8$ GeV and, from top to bottom, $m_H=125.2,125.158,125.157663$ GeV. 
We also fixed $\alpha_3(m_Z)=0.1184$. The shaded region is the range selected by our inflationary model: $10^{-3.4} \le V(\chi_0)^{1/4}/M\le 10^{-2.2}$.
The right panel is a magnification of the false vacuum region.}
\label{fig-Vmin}
\vskip .2cm
\end{figure}

As an example, in fig. \ref{fig-Vmin} we show the Higgs potential as a function of the Higgs field $\chi$,
by taking $m_t=171.8$ GeV and values of $m_H$ decreasing from $125.2$ down to $125.157663$ GeV from top to bottom. 
The plot shows that it is possible to have a second minimum at high energy (magnified in the right plot), 
in addition to the usual SM minimum at low energy. Having fixed $m_t$, this happens only for very specific values of $m_H$.
Increasing (decreasing) the top mass, the value of the second minimum $\chi_0/M$ increases (decreases), and larger (smaller) values of $m_H$ are required. 
The horizontal shaded band represents the range selected by the inflationary model
discussed in the previous section, namely $10^{-3.4} < V(\chi_0)^{1/4}/M<10^{-2.2}$.
The specific values considered in the plot are fine for our inflationary model.

As we are going to discuss, in order to have a sizable tunneling probability to the left side, the barrier must be very low, as is the case for 
 the middle curve in the right panel of fig. \ref{fig-Vmin}, obtained for $m_H =125.158$. 
For larger $m_H$ the potential has no false minimum, while for slightly smaller values of $m_H$ the second minimum becomes deeper and very soon negative.
In this case it is the SM minimum at low energy that becomes metastable and it could catastrophically tunnel to minus infinity. 
(The points in the $m_t-m_H$ plane corresponding to the transition from stability to metastability are shown in fig.\ref{fig-mtmh} as dashed lines;
the inner line is obtained when using the central value of $\alpha_3(m_Z)$, while the side ones represents its $1\sigma$ range.) 
For even smaller values of $m_H$ the tunneling probability increases so much to be inconsistent with the present age of the universe; the region 
of parameters is called instability region. 
For a discussion about the stability and metastability constraints of the Higgs potential see ref. \cite{hep-ph/0104016}.

The tunneling rate~\cite{Coleman} in our model is given by $\Gamma\simeq A\, e^{-B}$,
where $B$ is the Euclidean classical action of the bounce solution
which interpolates between $\chi_0$ and the value on the left side of
the barrier,  $\chi_T$, and $A$ is a dimension four quantity of the
order of the scale of the problem (GUT scale). We have numerically
computed $B$ for several potentials finding that only if the potential
is extremely shallow, such as the middle curve in the right panel of fig.\ref{fig-Vmin},  we
can obtain values of $B\lesssim{\cal O}(10^{2})$. If the potential
well is instead deeper as the bottom curve, the exponent $B$ becomes rapidly extremely large and $\Gamma$
becomes essentially zero. In this case the transition would never happen, or it would happen at a too low energy, 
leading to a too small reheating temperature.

As mentioned earlier note that the transition could happen also through a Hawking-Moss instanton (see for 
instance \cite{HenryTye:2008xu}), which 
is due the presence of a gravitational background. 
In the case in which this is the dominant process 
the whole scenario would be different because $\Gamma$ would also be time-dependent, since the Planck mass and $H$ are varying with time. 
Such a transition is basically due to quantum jumps of the fields if its mass (the second derivative of the potential at the false minimum) is smaller than $H$ (which is about $10^{11}-10^{12}$ GeV), and so it 
could be important for extremely shallow barriers and very high $H$ ({\it i.e.} very large values of $V(\chi_0)$). It may happen therefore that the scenario is viable only for low values of $V(\chi_0)$, which could translate on a further upper bound on the Higgs mass. In order to compare which instanton is the dominant one it is important for this analysis to have good control on the shape of the potential around the false minimum with a huge precision, of about 16 digits on $m_H$. We postpone this analysis for future work.
Note also that in the case in which such transition would not lead to sufficient inflation it is also possible to invoke an additional effect which would modify the Higgs field tunneling process, 
by considering a non-minimal coupling between the Higgs and gravity of the form $\xi \chi^2 R$. 
Since now $R$ is varying with time, this would introduce a time-dependence of the Higgs potential which 
could easily erase the potential barrier. We have in fact checked that a coupling $\xi$ of ${\cal O}(1)$ should be 
enough to erase the barrier, therefore opening another interesting possibility to implement inflation. However we leave also such a possibility for future work.

\section{The Higgs mass range}
\label{Results}

\begin{figure}[t!]
\includegraphics[width=11cm]{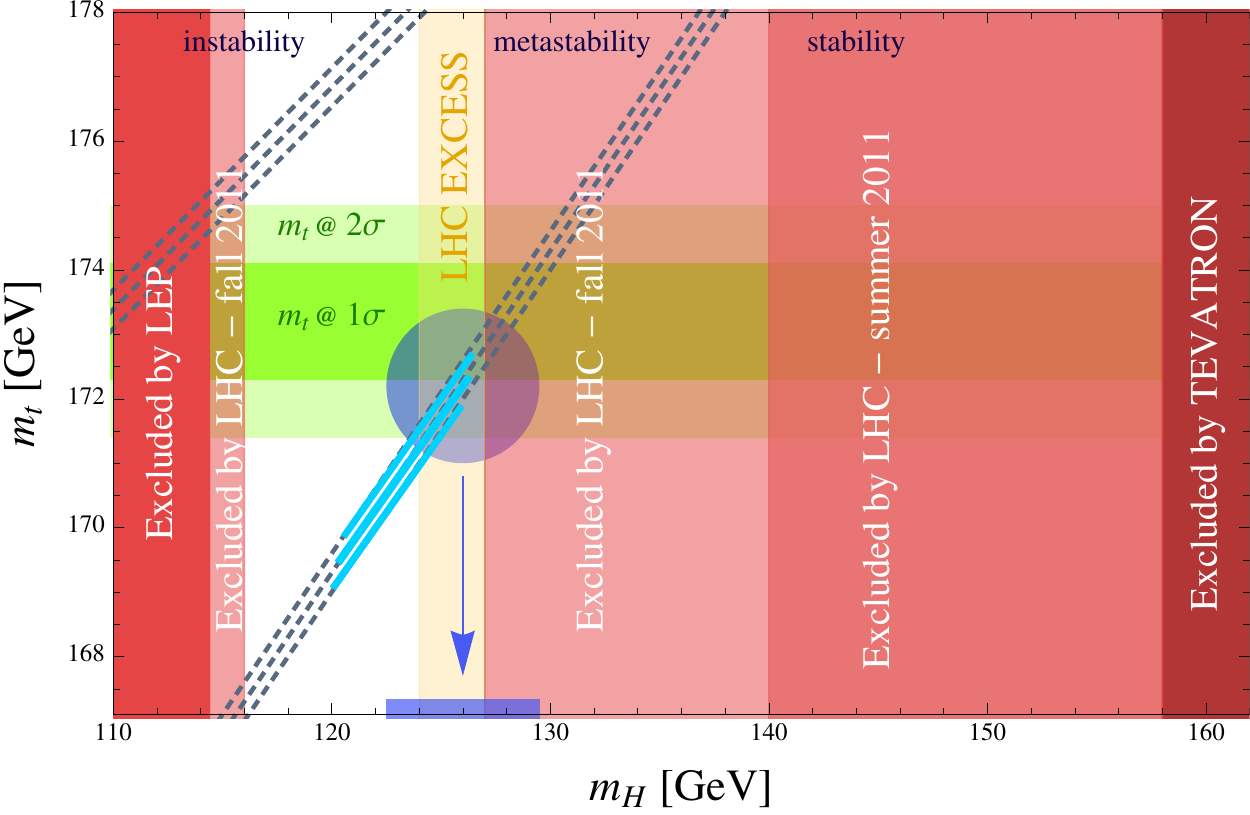}
\caption{ The (cyan) solid segments indicate the $m_t-m_H$ values compatible with a Higgs false minimum potential consistent with the requirement
from inflation,  fig.\ref{fig-Vmin}. The three segments show the uncertainty associated with $\alpha_3(m_Z)=0.1184 \pm 0.0007$  at $1\sigma$  (PDG 2011). 
The shaded horizontal bands are the $1\sigma$ and $2 \sigma$ ranges for $m_t$ from the GFitter analysis \cite{GFitter}. 
The central (blue) spot represents the region compatible with both SM false vacuum inflation and the experimental range of $m_t$:
this selects the narrow range $m_H=126.0\pm 3.5$ GeV, as pictorially represented via the arrow. 
The outer vertical shaded regions are the $m_H$ exclusion regions from direct searches at LEP, LHC and Tevatron. The central (yellow) vertical region
with $m_H=124-127$ GeV is the region of the LHC excess of events.
Dashed lines mark the values corresponding to the transition between stability, metastability and instability. }
\label{fig-mtmh}
\vskip .2cm
\end{figure}

As discussed, only with a restricted set of values of $m_H$ and $m_t$ it is possible for the false vacuum to be inside the band required by inflation.
These very particular values for $m_H$ and $m_t$ are displayed as segments in fig. \ref{fig-mtmh}. The upper (lower) values of $m_t-m_H$ in the band
correspond to the upper (lower) value of $V(\chi_0)^{1/4}$ allowed in our inflationary model, namely $1.5\times 10^{16}$ GeV ($9.7 \times 10^{14}$ GeV).
As can be seen in fig.\ref{fig-Vmin}, this in turn corresponds to a scalar-to-tensor ratio $r$ close to $0.1$ ($10^{-6}$).
The inner segment is obtained using the central value of $\alpha_3(m_Z)$, while the side ones mark its $1\sigma$ range. 

Our inflationary model works for a narrow band in the $m_t - m_H$  plane, which only partially 
overlaps with the top mass experimental range $m_t=173.2\pm0.9$ GeV,  provided by the recent (July 2011) global electroweak precision fits of the SM \cite{GFitter}.
We are then left with the upper part of the band in the $m_t-m_H$ plane, the one for which $V(\chi_0)^{1/4} \sim 10^{16}$ GeV and $r \sim 10^{-2}$.
Such region is emphasized with a  spot in fig.\ref{fig-mtmh} to
remind that the 2-loop RGE running has a theoretical uncertainty of $1$ GeV in $m_t$  and of $3$ GeV in $m_H$.

Consequently, the inflationary model based on the Higgs false minimum considered here gives a narrow prediction for the Higgs mass:
\be
m_H = (126.0 \pm 3.5)\,  {\rm GeV}\,\,.
\ee
This range is within the summer 2011 experimental window set by direct searches, namely the $115$ GeV lower bound set by LEP \cite{PDG} 
and the $140$ GeV  upper bound set by LHC \cite{HCP11}, which restricted the previous $155$ GeV bound set by Tevatron \cite{arXiv:1107.5518}.
Interestingly, this range is perfectly compatible with -- and even more precise than -- the recent global electroweak precision SM fit \cite{GFitter},
which gives $m_H=125^{+8}_{-10} $ GeV at $1\sigma$.

Hopefully the prediction of our inflationary model could be tested soon by LHC, 
so the model can  be experimentally supported or ruled out.
Actually, immediately after the first version of this preprint appeared on the arXiv, the ATLAS and CMS collaborations summarized the status of the Higgs boson searches
in a public seminar \cite{last}. The main conclusion is that, at $95\%$ CL, the SM Higgs boson mass range outside $116-130$ GeV is excluded by the ATLAS experiment, 
and the one outside $115-127$ GeV by CMS. It is extremely interesting that both experiments found some excess of events in the region between $124-127$ GeV,
mainly for the di-photon $H \rightarrow \gamma \gamma$ and four-lepton $H \rightarrow 4 l$ channels. Althought the significance of the excess is still too low,
these preliminary results from LHC are very exciting for the idea of false SM vacuum inflation.

If these preliminary experimental results were to be confirmed, it would be important to reduce the theoretical error which affects the determination
of the values of $m_t$ and $m_H$ allowing for the presence of a shallow SM false minimum.
The dominant source of the uncertainty in the RGE at present arises from the matching of the quartic Higgs coupling, known only at 1-loop. 
Such error is usually estimated by varying the matching scale in some (somewhat arbitrary) range. 
Choosing for example such range from about $125$ GeV (close to $m_H$) and about $175$ GeV (close to $m_t$) one finds that the value of $m_H$ 
leading to a shallow false minimum at the GUT scale changes by $1$ GeV.
In the literature one can find several different choices \cite{hep-ph/0104016,strumia2}  and perhaps a conservative 
error of $3$ GeV can be assigned on $m_H$,  and  of $1$ GeV on $m_t$.
This might overestimate the theoretical error, but in order to better understand it, one would need to know the 2-loop correction to the matching 
of the quartic Higgs coupling. Reasonably, one could expect that in this way the theoretical 
error on $m_H$ could be reduced down to $1$ GeV, which is comparable to the 
experimental precision on $m_H$ foreseen at LHC.


\section{Post-Tunneling evolution}
\label{Post}

When $\Gamma\simeq H^4$ many bubbles of the new phase quickly form, collide and rapidly give rise to a nearly  homogenous state 
with the Higgs field on the other side of the barrier at a value $\chi_F$ smaller than $\chi_0$, and some of its potential energy transformed into radiation. 
In this model, the fraction of energy converted into radiation is however very small, since the potential barrier has to be very shallow as can be seen in 
fig.~\ref{fig-Vmin}. So, the two values $\chi_0$ and $\chi_F$ between which the bounce solution interpolates are very close and the difference in potential energy 
$V(\chi_0)-V(\chi_F)$ before and after the transition is small compared to $V(\chi_0)$.  

It is well-known that a part of the energy density goes also into gravitational radiation through bubble collisions~\cite{Turner}, 
but for the same reason its energy density is going to be negligible compared to $V(\chi_0)$, unlike the case studied in~\cite{hep-ph/0511207} 
where it was assumed that the entire energy density would be converted into radiation and gravity waves through bubble collisions.

After this rapid thermalization, the Higgs field is free to roll down its potential, quickly reach zero and undergo rapid oscillations around zero. 
This happens at a rate faster than the expansion, since the quartic potential is steep and the term $V'(\chi)$ in the Klein-Gordon equation 
wins against the Hubble friction. During these oscillations the field is expected to rapidly convert all of its energy into particles, 
via perturbative and non-perturbative decays (similarly to what is described in~\cite{GarciaBellido:2008ab}). 
Since the Higgs couplings are large, it is expected that this process is very efficient, leading to reheating of the Universe, 
although probably through a very complicated sequence of processes. We may however assume that roughly all the energy density 
$V(\chi_F)\simeq V(\chi_0)$ is converted into a bath of SM particles: equating  the energy density of the produced radiation 
to the initial potential energy we get an estimate for the reheating temperature $g_{*} T_{RH}^4\simeq V(\chi_0)$, 
where $g_*$ is the number of degrees of freedom of the SM,  $g_*=106.75$. 
After this, the Universe cools down as usual and finally the Higgs field settles down to its present electroweak vacuum expectation value $v \sim 246 \, {\rm GeV}$.

Apart from the evolution of the Higgs field, it is also necessary to follow the evolution of the $\Phi$ field. 
In fact, the $\Phi$ field has to satisfy the fifth  force constraints at late times, due to the fact that an additional light scalar 
can mediate a long-range force between matter bodies and there are strong constraints about this, especially from the solar system~\cite{reviewPPN}.  
Those constraints can be satisfied if the following quantity~\cite{espositofarese}
\be
\alpha \equiv \frac{d \ln A}{d \Phi}  \,\,\, , \,\,\, A\equiv f^{-1/2}
\ee
is small today, $\alpha^2\lesssim 2\times 10^{-4}$~\cite{reviewPPN}, or if the field is massive enough $m_{\Phi}\gtrsim 1 eV$, so that it does not mediate a long-range force~\cite{reviewPPN}. 
Moreover, it is necessary that at least after Big Bang Nucleosynthesis (BBN) the field $\Phi$ do not evolve significantly, 
because the value of the Planck mass (set by $f(\phi)$) is constrained to be close to its present value~\cite{reviewPPN}.

It is difficult to predict the evolution of the field $\Phi$ during the oscillations of the Higgs, 
because this would require to compute the equation of state $w=p/\rho$ where $p$ is the pressure and $\rho$ is the 
energy density of the total amount of matter contained in the Universe. 
In the absence of dissipation and in the approximation in which the Higgs field has just a quartic potential this would be possible, 
since it is well-known that averaging oscillations on a quartic potential leads to an equation of state of $w=1/3$. 
Moreover for the radiation  produced the equation of state is also $w=1/3$. 
However the fact that the energy gets dissipated decreases the kinetic energy of the field $\Phi$, leading probably to a $w$ slightly smaller than $1/3$. 
Even when the oscillations are completed the equation of state of radiation is not exactly $w=1/3$ but it is slighly smaller because of quantum corrections which break the conformal invariance, due again to the running of couplings (mostly QCD corrections~\cite{Peloso}). 
The value $w=1/3$ is critical because the field is driven to large values if $w<1/3$ or to smaller values if $w>1/3$, 
and so we would conclude that the field $\Phi$ after tunneling stays at a large value, close to the value that it takes at the tunneling epoch, $\Phi_F$, or slightly larger.

On the other hand, in order for our model to be predictive we need to identify $M$ with the present value of the Planck mass $M_{Pl}$, otherwise we would not be able 
to tell what is the value of the scale $V^{1/4}(\chi)$ needed from fig.~\ref{fig-Vmin}, 
which leads to our prediction on the Higgs mass. 
This can be achieved if the post-tunneling evolution of $\Phi$ drives it back to zero, and in this case it automatically follows that the solar-system constraints are satisfied, since $\alpha$ is very close to zero. 
As we have just discussed this could happen if $w>1/3$ or, alternatively, upon introduction of a potential $U(\phi)$ in~(\ref{azione}).  
Since we do not want to alter the SM significantly, it seems that the latter option is to be taken.

 In this case, we can analyze the evolution again in the Einstein frame, remembering that a potential term $U(\phi)$ becomes 
 now a potential $\bar{U}(\Phi)=U(\phi(\Phi))/f(\phi(\Phi))^2$. Since we want $\Phi$ to go to zero, we have to require 
 that $U(\phi)$ is a function which grows more rapidly than $f^2(\phi)$ for large $\phi$. 
 On the other hand we also want $U$ to be negligible before tunneling, so the choice of $U$ requires some care, 
 but it can be shown that such functions can be constructed. 
 
 An additional relevant issue is that after tunneling the field $\Phi$ may lead to some additional inflation, because the potential $U$ effectively can 
 introduce a slow-roll phase, since the field is at values $\phi\gg M$, similarly to what happens in chaotic inflation models~\cite{Linde}.
This would shift the needed number of e-folds $\bar{N}$ by some model-dependent number $\Delta \bar{N}$, which has to be not too large 
in order not to erase the predictions of our model discussed in the previous section.

Although these are relevant and interesting issues, they are however very model-dependent, and it is sufficient to say for our purposes 
that a mechanism to drive back the field $\Phi$ to zero has to be implemented, 
most likely using a potential $U(\phi)$, and this has to be done without introducing too many e-folds of an additional inflationary phase in order for 
our predictions to be valid. 
Under the assumption that $\Delta {\bar N}$ is zero or negligible we can compute the value of $\bar N$ which corresponds to our present
horizon as follows.

First of all, let us compute when a particular comoving scale $L$ went outside the horizon during inflation. We count the number of e-folds
starting from the end of exponential inflation (whose scale factor we call $a_E$), going backwards in time. In general a scale $L$ leaves
the horizon at some e-folding number $\bar{{N}}$ if:
\be
L \left(\frac{T_0}{T_{\rm{RH}}} \right)\left(\frac{\bar{a}_{E}}{\bar{a}_{RH}} \right) e^{- \bar{{N}}}=\bar{H}_{I}^{-1} \, .
\ee
The reheating temperature is given by $T_{RH}^4\simeq V(\chi_0)/g_{*} $ and
the redshift during the power-law phase is given by
$\bar{a}_{E}/\bar{a}_{RH}=(\bar{t}_{E}/\bar{t}_{RH})^{3/4} \simeq( \bar{\Gamma}^{1/4} / \bar{H}_I)^{3/4}$.
Here we have assumed that the transition between the exponential and the power-law phase is very quick and this is true for most
functions, $f(\phi)$. Now, the largest scale observed today is the
horizon scale, which leads to:
\be
{\bar N} \lesssim 60\,\,,
\ee
where we have used $V(\chi_0)^{1/4} \approx 10^{-2.4} M$, $L=3000$ Mpc$/h$, $h\approx 0.7$ and we have taken 
the extreme value ${\bar \Gamma}^{1/4} \approx {\bar H}_I$.

Also, if the field is driven to zero, it would start oscillating around the zero and it could overclose the Universe~\cite{moduli}, 
unless it can decay. 
Since we do not want in principle to introduce new couplings, this can be achieved either by decay into $\Phi$ quanta, 
through self-interactions, or via decay into gravitons. The latter can be estimated to have a decay rate of the order of 
$m_{\Phi}^3/M_{Pl}^2$, and the decay has to happen before BBN, which has a temperature of ${\cal O }({\rm MeV})$. 
This leads to the requirement that the decay rate is larger than the expansion rate: $m^3_{\Phi}/M^2_{Pl}\gtrsim{\rm MeV}^2/M_{Pl}$ 
which leads to $m_{\Phi}\gtrsim 10^4 {\rm GeV}$. 

\subsection{Further issues}

Finally, let us comment on the possibility that other cosmological puzzles have to be solved in this minimalistic model with SM and tensor-scalar gravity. 
It is well-known for example that other ingredients need to be introduced beyond the SM 
in order to satisfy cosmological observations, such as baryogenesis and dark matter.

For what concerns baryogenesis it is interesting to note that the tunneling event provides an out-of-equilibrium event in the early Universe 
which could be used for generating the baryon asymmetry. This would require only the addition of some source of CP-violation. 
While in the usual SM electroweak phase transition the CP-violation is considered to be too small  
 it would need to be investigated if this remains true in such a scenario.  In any case, it is always possible to add right-handed neutrinos 
 and achieve a new-source of CP-violation in the lepton sector of the SM without changing significantly 
 the running of $\lambda$ (which depends mostly on the gauge and top Yukawa couplings) and therefore without affecting at all our scenario. So baryogenesis could proceed in a new non-trivial way during reheating and oscillations of the Higgs field, or it could be also obtained via the usual leptogenesis scenario~\cite{Yanagida}, where right-handed neutrinos are thermally produced and decay out-of-equilibrium, since the reheating temperature in this model can be large enough.
 
For the dark matter problem many solutions could be possibly incorporated in our model. For instance, there could simply be an additional 
weakly-interacting stable particle, which again would not change significantly the running of the SM.
Alternatively, it could be worth considering the hypothesis that quanta of the field $\Phi$ itself are left as a remnant of the post-tunneling 
evolution of $\Phi$. Finally, in the same spirit of the proposal of the present paper, it is possible also to add another particle completely 
decoupled from the SM, which could perhaps be produced gravitationally or via decay of the $\Phi$ field during or after inflation.

\section{Exponential inflation with quadratic and higher oder terms} 
\label{n>2} \label{General}

\subsection{Effect of $\beta$}

We study now the inclusion of the quadratic coupling $\beta$ in eq.(\ref{eq-ftot}) and in order to do this we focus on the case $n=4$, 
so that 
\begin{equation}
 f(\phi)\simeq 1+\beta \left(\frac{\phi}{M}\right)^2 + \gamma_4  \left(\frac{\phi}{M}\right)^4 \,\,.
\end{equation}
We study the evolution equations both analytically and numerically.

\begin{figure}[h!]
\includegraphics[width=7 cm]{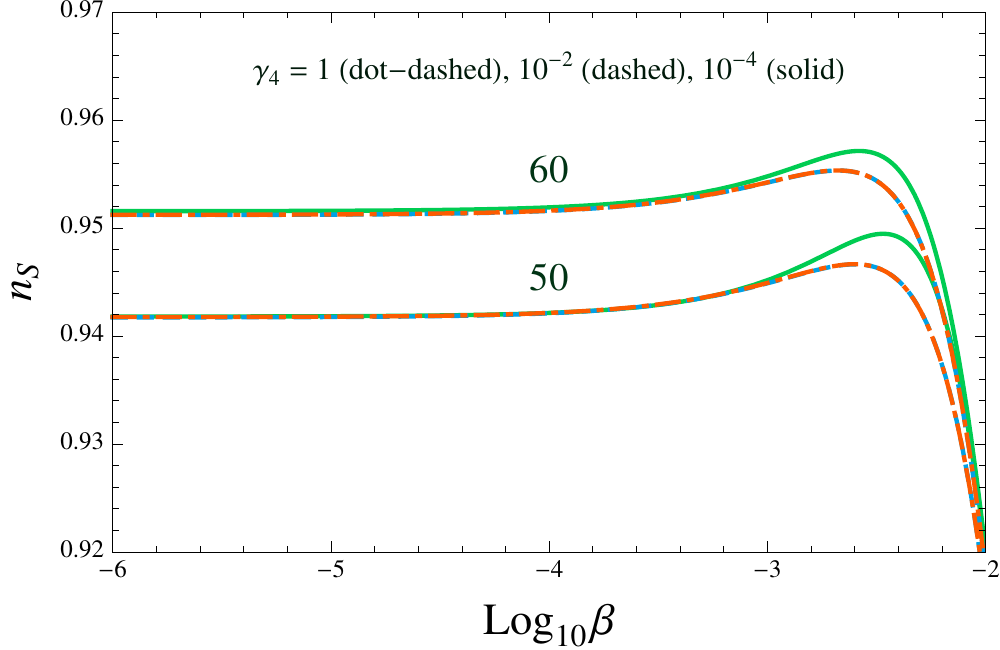} \,\,\,\,\, \includegraphics[width=7 cm]{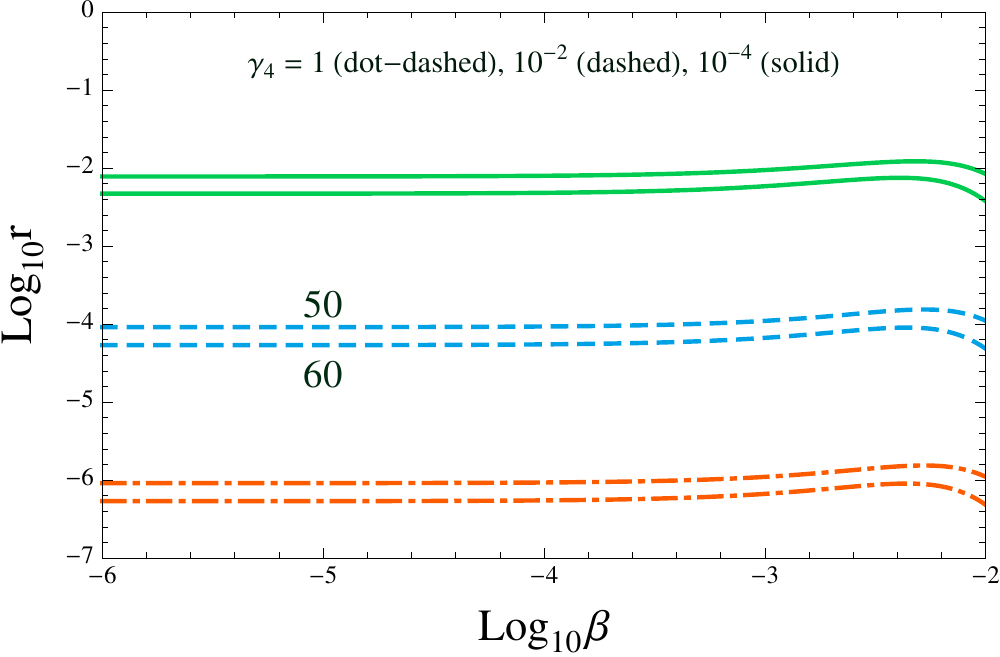}
\caption{Analytical calculation of $n_S$ and $r$ as a function of $\beta$ for $n=4$. Values of $\gamma_4$ and $\bar N$ are indicated in the plots. }
\label{fig-beta}
\vskip .2cm
\end{figure}

For the analytical calculation, we work directly in the Einstein frame, computing $\Phi(\phi)$ by solving the differential equation $d\Phi= d\phi \sqrt{K(\phi)}$
via a power series expansion in the parameter $\beta$. We thus obtain an expression for the potential $\bar{V}(\Phi)$ which generalizes the one in eq.(\ref{eq-Vb})
\begin{equation}
\bar V\left(\Phi\right) =V(\chi_0) \left[   1 -2 \beta \left( \frac{ \Phi}{M} \right)^2    + \left( \frac{13}{3} \beta^2  +16 \beta^3-2 \gamma_4 \right)   \left( \frac{ \Phi}{M} \right)^4 + ...  \right]\,\,.
\end{equation} 
The number of e-folds is calculated using the first approximation of eq.(\ref{Nphi}). By selecting a value for the number of e-folds $\bar N$, one obtains the
associated predictions for $n_S$ and $r$, as displayed in fig.\ref{fig-beta}. 
As already anticipated, as far as $\beta \lesssim 5\times 10^{-4}$, the prediction for $n_S$ does not change significantly
with respect to the case in which $\beta=0$. This can be seen from the left plot of fig. \ref{fig-beta}. For higher values of $\beta$ there is a slight
increase in $n_S$, which then falls down in the experimentally excluded region at $\beta \approx 10^{-2}$. 
The dependence of $r$ on $\beta$ is less pronounced, see the right plot of fig. \ref{fig-beta}. When $\beta \gtrsim 10^{-2}$, $r$ falls down rapidly.

\begin{figure}[h!]
\includegraphics[width=7 cm]{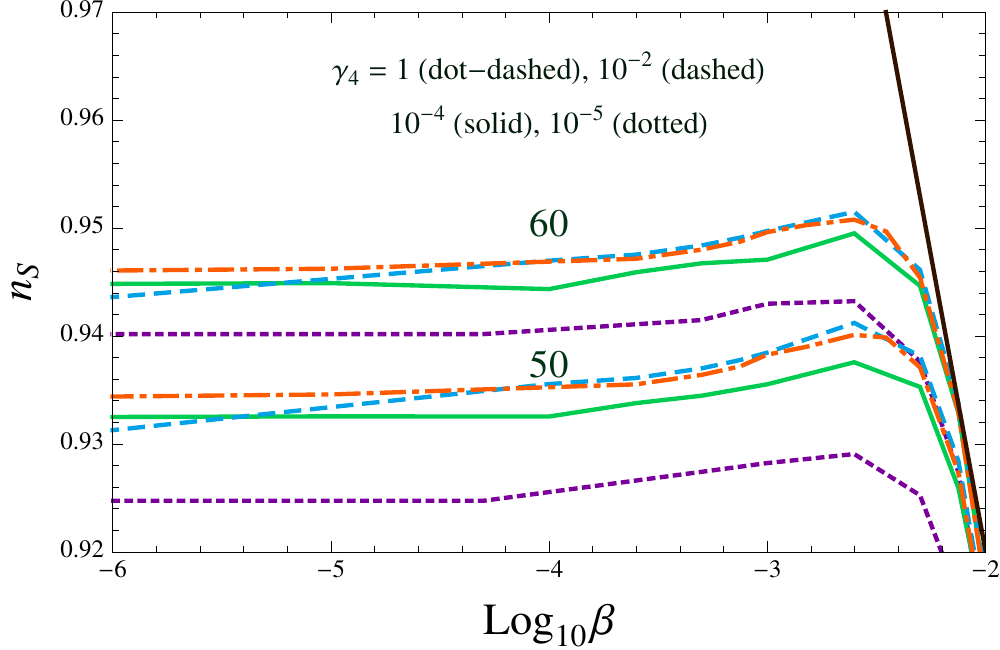} \,\,\,\,\, \includegraphics[width=7 cm]{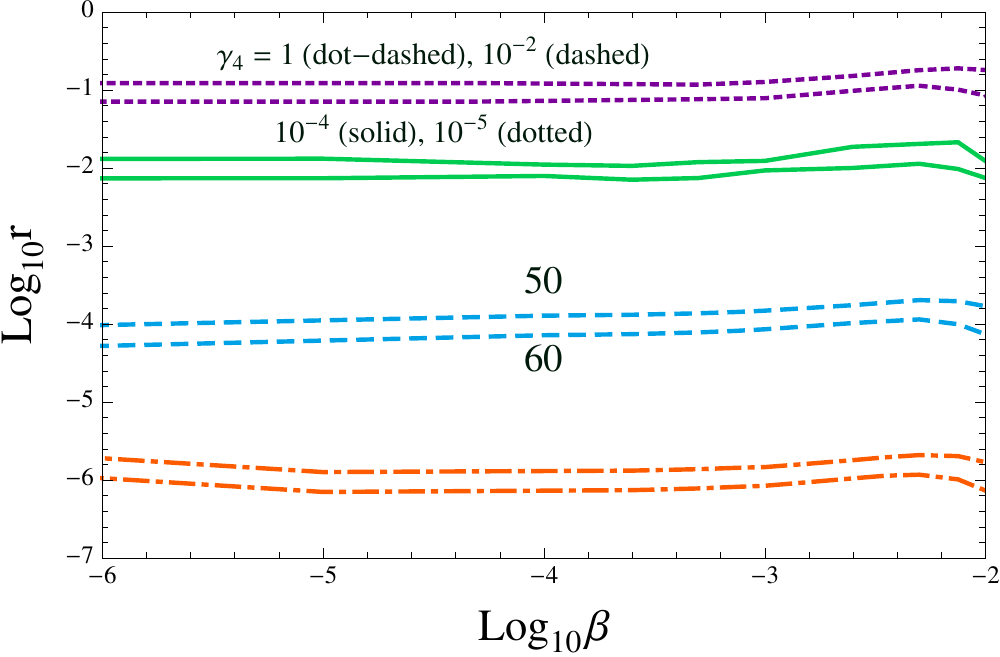}
\caption{Numerical calculation of $n_S$ and $r$ as a function of $\beta$ for $n=4$. Values of $\gamma_4$ and $\bar N$ are indicated in the plots. 
The dark solid nearly vertical line is $n_S=1-8 \beta$, which is the behavior when $\beta$ dominates as derived in \cite{hep-ph/0511207}. }
\label{fig-beta-num}
\vskip .2cm
\end{figure}

Also for the numerical analysis we work directly in the Einstein frame, computing $\Phi(\phi)$ by solving the differential equation $d\Phi= d\phi \sqrt{K(\phi)}$, 
which gives us the potential $\bar{V}(\Phi)$ of eq.(\ref{eq-Vb0}). 
We solve then the equations of motion for $\Phi$ and the Friedmann equation in order to find the evolution of the scale factor $\bar{a}(t)$.
The number of e-folds is computed as $\bar N(\bar{t})=\int_{{\bar{t}_F}}^{\bar{t}} \bar{H}(t) dt$, where ${\bar{t}_F}$ corresponds to the end of inflation, 
and the slow roll parameters $\epsilon_{\bar N}$ and $\eta_{\bar N}$ have been computed as usual by computing the first and second derivatives of $\bar{V}$. 
The result for $n_S$ is given in fig.\ref{fig-beta-num} for $\bar N=50$ and $\bar N=60$. 
Here $t_F$  is here defined as the time in which $\ddot{a}$ becomes negative. Note that close to $t_F$ the slow-roll parameters are already ${\cal O }(1)$.
This leads to a slight difference compared to the approximate expression of eq.(\ref{Nphi}) where the end of inflation has been defined there as the moment
 when either $\epsilon$ or $\eta$ are equal to $1$, and which is valid only when $\epsilon\ll 1$: all this leads to a shift in $\bar N$ of about 5. 
 Note also that  if $\Gamma$ is large enough the tunneling event could happen {\it before} the transition to the decelerated phase, 
 because $H$ is already decreasing before that the asymptotic behavior $\bar{a}\propto \bar{t}^{3/4}$ is reached. In this case there can be another shift in $\bar{N}$, 
 which makes the spectral index slightly higher and $r$ slightly smaller: for instance for the case $\gamma=10^{-5}$ and $\beta=10^{-2}$ the numerical calculation 
 can make $n_S$ increase by $0.003$ and $r$ decrease by $0.03$ in the numerical result.
Anyway remind that $\bar N$ is subject to other uncertainties due to post-inflationary evolution, as discussed in the previous section.


\subsection{Higher order terms}
We derive in this section the predictions for the case in which
\be
f(\phi)\simeq \gamma_n \left( \frac{\phi}{M}\right)^n
\ee
where $n=4,6,8,...$, further elaborating some results already derived in \cite{hep-ph/0511207}.
Again, for $\phi\ll M$, the canonical scalar field variable $\Phi$ and $\phi$ almost coincide ($K\approx 1$) and to lowest order in $\Phi/M$ we get:
\be
\bar{V}(\Phi)= V(\chi_0)\left[1-2\gamma_n\left(\frac{\Phi}{M}\right)^n\right] \, .
\ee
In the slow-roll approximation, 
\begin{eqnarray}
\epsilon (\Phi)
\approx 2 \, \gamma_n^2 \, n^2 \, \left(\frac{\Phi}{M} \right)^{2(n-1)} \,\, , \,\,\,\,\,
\eta (\Phi) 
\approx -2 \, \gamma_n \, n(n-1) \, \left(\frac{\Phi}{M} \right)^{n-2} 
\label{eq-epseta} \, .
\end{eqnarray}

We call $\Phi_f$ the field $\Phi$ at the end of inflation, for definiteness say when $\epsilon$ or $\eta$ become equal to one. 
For $\gamma_n \lesssim 1/(2^{n/2} n (n-1)^{n-1})$, one has $1=\epsilon(\Phi_f) \gtrsim \eta(\Phi_f)$, otherwise $\epsilon(\Phi_f) \lesssim \eta(\Phi_f)=1$. 
The number of e-foldings as a function of $\Phi$ is given by
\be
{N} (\Phi)  \approx \frac{1}{M} \int_{\Phi}^{\Phi_f} d{\Phi'}\, \frac{1}{\sqrt{2 \, \epsilon(\Phi')}}
\approx \frac{1}{2 n (n-2) \gamma_n }  \left(  \frac{1}{ \left(\frac{\Phi}{M} \right)^{n-2}} -\frac{1}{ \left(\frac{\Phi_f}{M} \right)^{n-2}}  \right)  \,\,,
\ee
which can be rewritten as 
\be
\left(\frac{\Phi_f}{M} \right)^{n-2} \approx \frac{1}{2 n (n-2) \gamma_n { N}(\Phi) + C_{\gamma_n}} 
\label{eq-Cn}
\ee
where $C_{\gamma_n}\approx (\sqrt{2} n \gamma_n)^{\frac{n-2}{n-1}}$ for $\gamma_n \lesssim 1/(2^{n/2} n (n-1)^{n-1})$, and
$C_{\gamma_n}\approx 2 n (n-1) \gamma_n$ otherwise.

\begin{figure}[t!]
\includegraphics[width=7 cm]{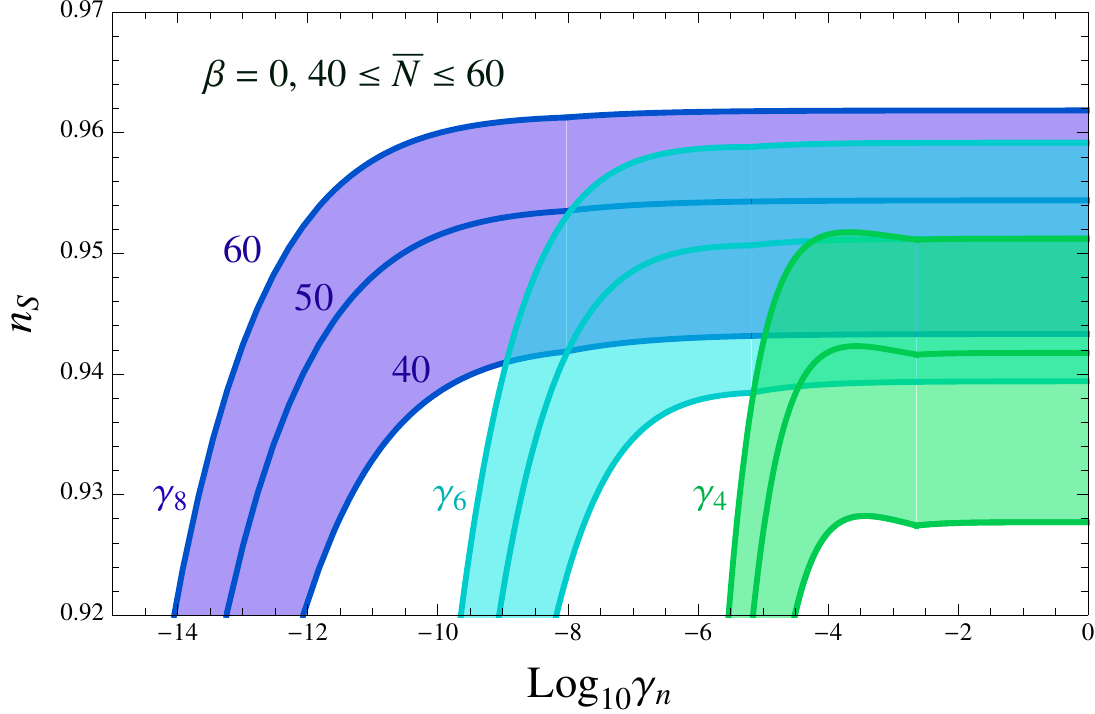} \,\,\,\, \includegraphics[width=7 cm]{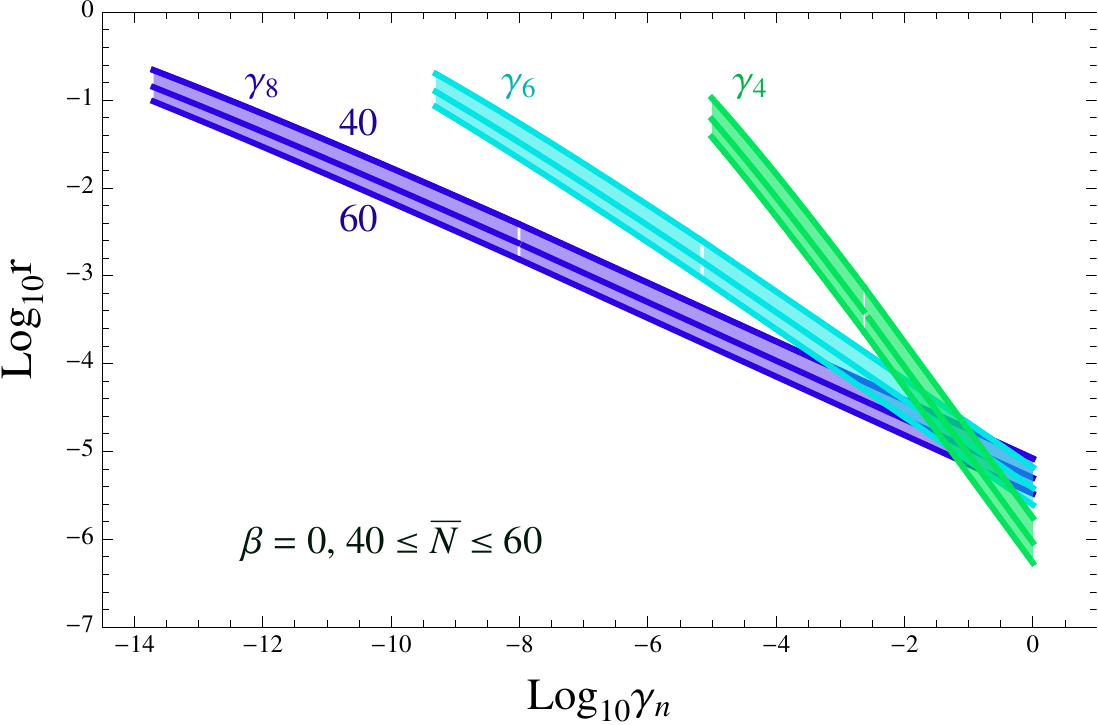}
\caption{$n_S$ and $r$ as a function of $\gamma_n$ for $n=4,6,8$ and  $\beta=0$. The shaded regions are obtained
by varying $\bar N$ in the representative interval $40-60$.}
\label{fig-nsrgn}
\vskip .2cm
\end{figure}

This expression can be substituted again in eqs. (\ref{eq-epseta}) in order to obtain observable quantities as
the tensor to scalar ratio $r  = 16 \epsilon_{\bar N}$ and the scalar spectral index $n_S=1 -6 \epsilon_{\bar N} +2 \eta_{\bar N}$, 
for a fixed number of e-folds $\bar N=N(\Phi)$ and as a function of $\gamma_n$, as shown in fig.\ref{fig-nsrgn} for $n=4,6,8$. 
One can see that, when $\gamma_n$ is large enough so that $C_{\gamma_n}$ is negligible in the denominator of
eq.(\ref{eq-Cn}), $n_S$ becomes nearly constant and it is approximately given by:
\be
n_S \approx 1- \frac{n-1}{n-2}\frac{2}{\bar N}  \,\,.
\ee
Hence, $n_S$ increases with $n$ and eventually reaches the maximum value $n_S \approx 1-2/{\bar N}$, which is close to $0.967$ for $\bar N=60$.
The prediction for $n_S$ is well inside its present observational range for reasonable values of $\bar N$.
On the other hand, the tensor-to-scalar ratio is shown in the right plot of fig.\ref{fig-nsrgn} for $n=4,6,8$ and can be approximated by
\be
r \approx 32\, n^2 \, \gamma_n^2 \, \frac{1}{  (2 n (n-2) \gamma_n {\bar N})^{\frac{2(n-1)}{n-2}}} \,\,.
\ee
Observationally $r\lesssim 0.1$, which implies the lower bounds:
$\gamma_4\gtrsim 10^{-5}$,  $\gamma_6\gtrsim 10^{-9}$, $\gamma_8\gtrsim 10^{-13}$. 

\begin{figure}[t!]
\includegraphics[width=10.1 cm]{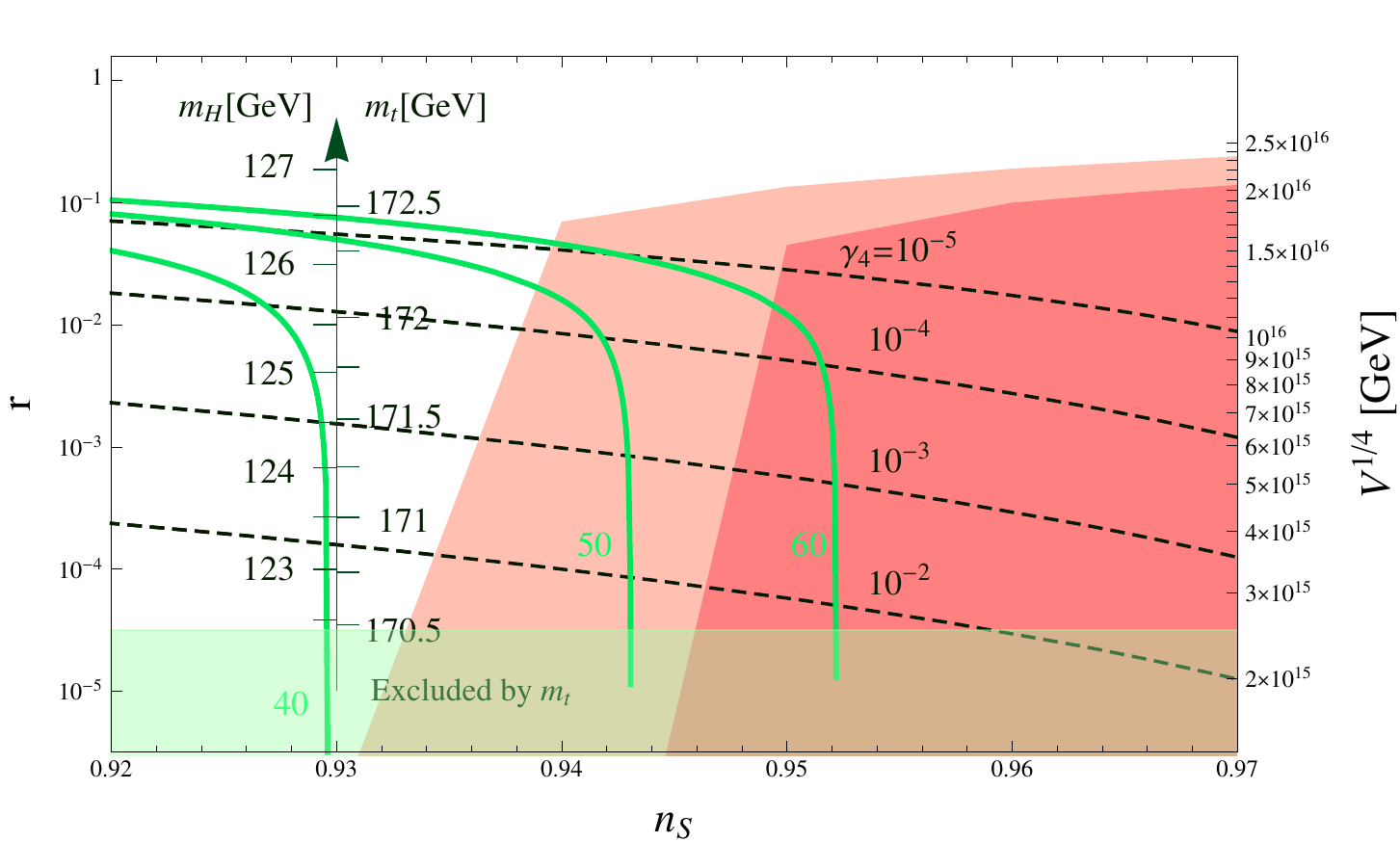} \vskip.5cm \includegraphics[width=10.1 cm]{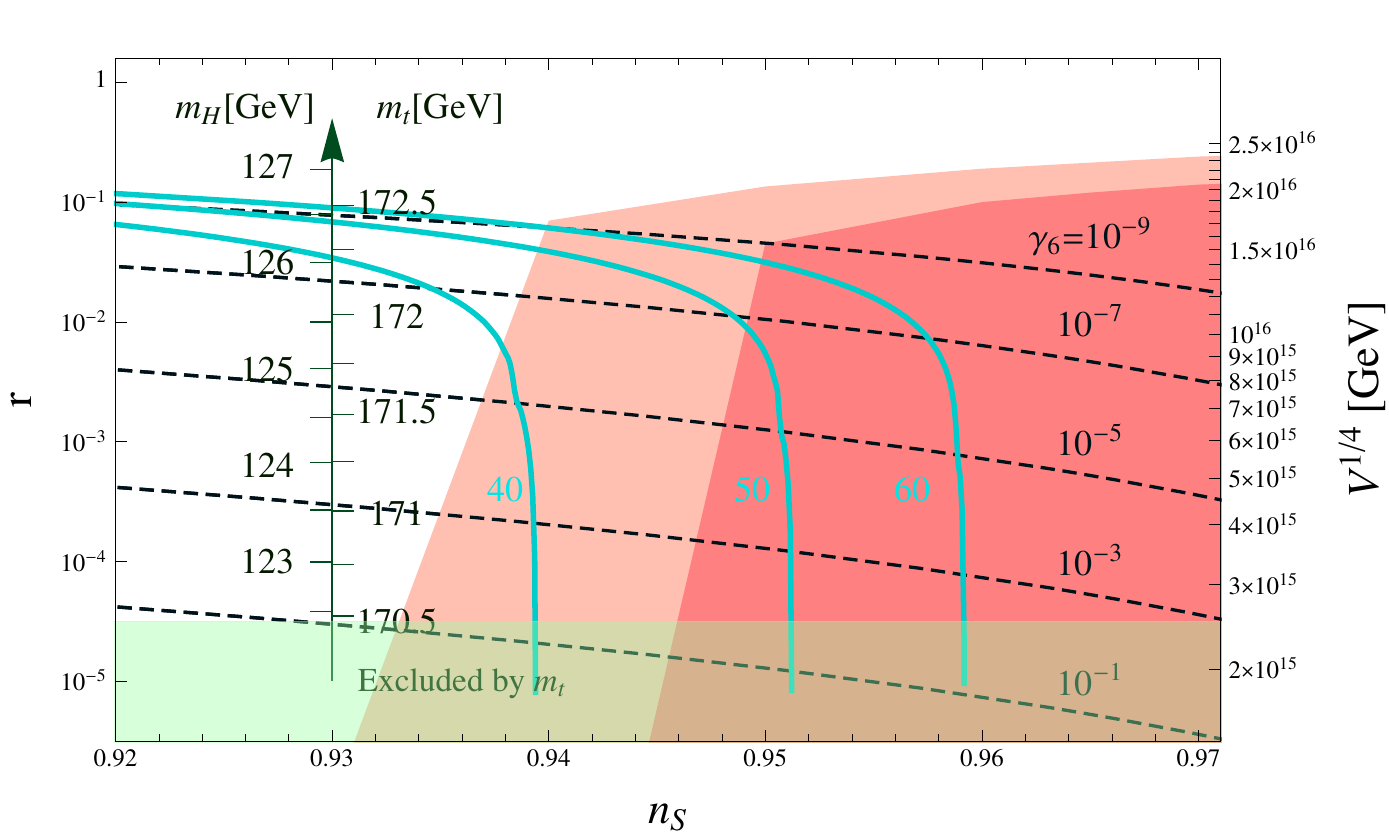} \vskip.5cm \includegraphics[width=10.1 cm]{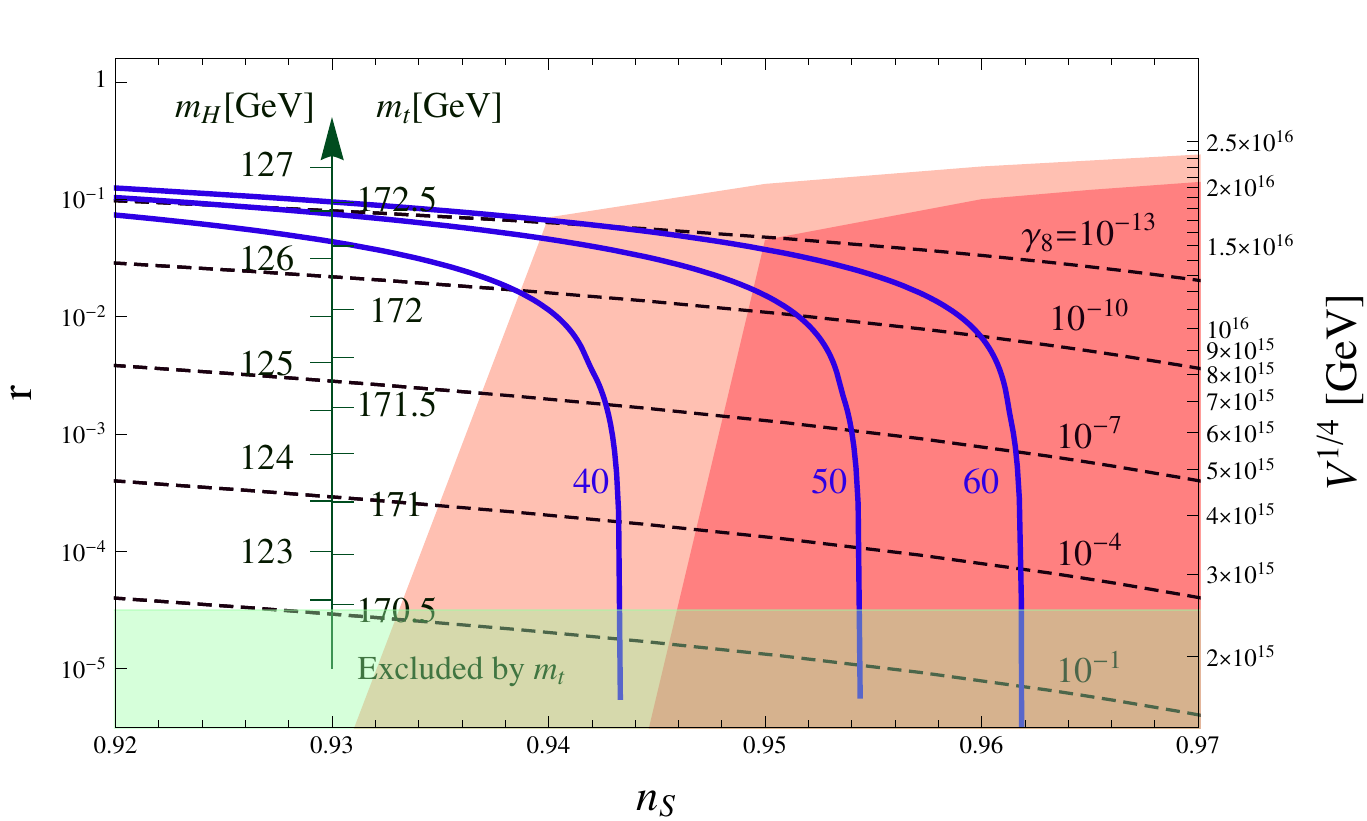}
\caption{Connection between $r$, $n_S$, $\gamma_n$, $m_H$, $m_t$ and $V(\chi_0)^{1/4}$ for $n=4,6,8$ from top to bottom.  
The solid lines are such that $\bar N =40,50, 60$. The central shaded (red) regions are to the $1$ and $2\sigma$ ranges 
allowed experimentally for $r$ and $n_S$ \cite{Komatsu:2010fb}. The horizontal bottom region is excluded by the $2\sigma$ lower bound on $m_t$ \cite{GFitter}.
It is understood that there is an uncertainty of about $3$ GeV in the ticks for $m_H$ and of about $1$ GeV in those for $m_t$.}
\label{fig-rns}
\vskip .2cm
\end{figure}

In fig.\ref{fig-rns} we show the (dashed) curves of constant $\gamma_n$ in the plane $r-n_S$, for $n=4,6,8$ from top to bottom. The plot also
show the (solid) curves of constant $\bar N$, considering as reference values $\bar N=40,50,60$.
Due to the proportionality between $r$ and $V^{1/4}(\chi_0)$, it also possible to display for each value of $r$ the values of the Higgs and top masses
giving rise to a shallow false minimum; this is represented graphically via the vertical arrow.  The lower $2\sigma$ bound on $m_t$ is about $171.5$ GeV  \cite{GFitter} 
but, as already explained, in the determination of the shallow minimum there is theoretical uncertainty of about $3$ GeV in $m_H$ and of about $1$ GeV in $m_t$;
we have thus displayed the exclusion region for $m_{t}$ by means of the shaded region below $170.5$ GeV. 
The plots show that the three models with $n=4,6,8$ are well compatible with the present observed values of $r$ and $n_S$ \cite{Komatsu:2010fb} at $1$ and $2 \sigma$,
displayed via the shaded red regions. They also show that, due to the lower bound on $m_t$, in these models $r$ should be found above roughly $10^{-4}.$

\section{Conclusions}
\label{Conclusions}

We proposed to exploit the possible presence of a false minimum in the SM Higgs potential at very high-energies, $10^{15}-10^{17}$ GeV,
to provide a large amount of potential energy which can drive primordial inflation. 
In the framework of scalar-tensor theories of gravity, a graceful exit from inflation can be achieved through tunneling of the Higgs field and subsequent relaxation
down to its present vacuum expectation value $v\sim 246$ GeV.

Requiring the amplitude and spectral index of cosmological density perturbations from inflation
and the top quark mass to be compatible with observations, 
we showed that this possibility is realized only within a small region of values for the Higgs mass, see fig.\ref{fig-mtmh},
leading to the prediction
$$
m_H = (126.0 \pm 3.5)\,  {\rm GeV}\,\,,
$$
where the error is mostly due to the present theoretical uncertainty of the 2-loop RGE. 
This prediction can be tested soon by LHC, in particular for the decay mode $H \rightarrow \gamma \gamma $.
The inflationary model proposed here could thus meet experimental support or be ruled out.
It is exciting that preliminary results from LHC \cite{last} show an excess of events in the range $124-127$ GeV.

If these results will be confirmed, further checks of the model will be offered by better determinations of the scalar spectral index $n_S$,
predicted in this model to be within $0.93-0.96$. 
Using the constraints coming from the top quark mass,
the scalar-to-tensor ratio $r$ in this model is constrained to be within $ 10^{-4}\lesssim r \lesssim10^{-1}$.
Actually, it can be shown \cite{Masina:2011un} that any model in which the false vacuum is very shallow, the relation between the potential 
at the false minimum and the amplitude of perturbations, eq. (\ref{eq-Dr}), implies $r \gtrsim 10^{-4}$ .
With forthcoming more precise cosmological measurements, such as the Planck satellite mission, one can test the region of large values of $r$,
while improving the top quark mass measurement can further constrain $r$ from below. 

Moreover, we showed that our scenario, and in particular the prediction for $n_S$, can be obtained in a wide class of scalar-tensor theories.
In particular, higher dimensional operators can be safely present.

As a completely general remark, we point out that 
discovering a Higgs with mass close to $126$ GeV is a very suggestive hint in favor of the existence of a {\it false minimum} in the SM Higgs potential
at energies close to $10^{16}$ GeV, which could be {\it the starting point for inflation in our Universe.} Indeed, it would lead to a period of exponential expansion
producing  density perturbations with the right amplitude. 
We have shown here that in a scalar-tensor theory of gravity this inflationary stage can end, allowing the Higgs field to tunnel out of the false minimum 
and subsequently to relax down to its present electroweak scale value.

\vskip .7cm
{\bf Acknowledgements} We thank Tirthabir Biswas for discussions in the early stages of this project and Pietro Dalpiaz for interesting suggestions. 
We also thank Jaume Garriga and Giovanni Villadoro for useful comments.


\end{document}